\documentclass[aip,cha,reprint]{revtex4-2}

\usepackage{amsthm,amsfonts,amssymb,amsmath,dsfont}
\usepackage[pdftex]{graphicx,graphics,color}
\usepackage{dcolumn}
\usepackage{bm}
\usepackage{color,colortbl}
\usepackage[T1]{fontenc}
\usepackage[latin1]{inputenc}
\usepackage{lmodern}
\usepackage[english]{babel}
\usepackage[colorlinks=true, linkcolor=black, citecolor=black, 
	anchorcolor=black, filecolor=black, urlcolor=black]{hyperref}

\begin{document}

\newcommand{\bra}{\langle}
\newcommand{\ket}{\rangle}
\newcommand{\del}{\partial}
\newcommand{\dg}{^{\dag}}
\newcommand{\cg}{^{*}}
\newcommand{\T}{^{T}}
\newcommand{\vep}{\varepsilon}
\newcommand{\suml}{\sum\limits}
\newcommand{\prodl}{\prod\limits}
\newcommand{\intl}{\int\limits}
\newcommand{\til}{\tilde}
\newcommand{\mcl}{\mathcal}
\newcommand{\mfk}{\mathfrak}
\newcommand{\mds}{\mathds}
\newcommand{\mbb}{\mathbb}
\newcommand{\mrm}{\mathrm}
\newcommand{\mnl}{\mathnormal}
\newcommand{\ds}{\displaystyle}
\newcommand{\rmi}{\mathrm{i}}
\newcommand{\rme}{\mathrm{e}}
\newcommand{\rmd}{\mathrm{d}}
\newcommand{\rmD}{\mathrm{D}}
\newcommand{\vphi}{\varphi}
\newcommand{\malt}{\mathpzc}
\renewcommand{\-}{\,-}

\title[Slippery-Sticky Transition of Interfacial Fluid Slip]
{Slippery-Sticky Transition of Interfacial Fluid Slip}

\author{Thiago F. Viscondi}
	\email{viscondi@usp.br}
	\affiliation{Escola Polit\'ecnica, University of S\~ao Paulo, S\~ao Paulo 05508-030, Brazil}
\author{Adriano Grigolo}
	\affiliation{Escola Polit\'ecnica, University of S\~ao Paulo, S\~ao Paulo 05508-030, Brazil}
\author{Iber\^e L. Caldas}
	\affiliation{Institute of Physics, University of S\~ao Paulo, S\~ao Paulo 05508-090, Brazil}
\author{Julio R. Meneghini}
  \affiliation{Escola Polit\'ecnica, University of S\~ao Paulo, S\~ao Paulo 05508-030, Brazil}

\date{\today}

\begin{abstract}
The influence of temperature on interfacial fluid slip, as 
measured by molecular-dynamics simulations of a Couette flow 
comprising a Lennard-Jones fluid and rigid crystalline walls, 
is examined as a function of the fluid-solid interaction strength. 
Two different types of thermal behavior are observed, namely, the 
slippery and sticky cases. The first is characterized by a steep 
and unlimited increase of the slip length at low temperatures, 
while the second presents a vanishing slip length in this regime. 
As the temperature increases in relation to a characteristic value,
both cases converge to finite slip lengths. A recently proposed 
analytical model is found to well describe both thermal behaviors, 
also predicting the slippery-sticky transition that occurs at a 
critical value of the fluid-solid interaction parameter, for which, 
according to the model, fluid particles experience a smooth average 
energy landscape at the interface.
\end{abstract}

\maketitle


\section{Introduction}
\label{sec:Introduction}

Fluid-solid interaction is a fundamental topic in several 
areas of science and engineering.\cite{lyklema1995fundamentals, 
wandelt2002solid, wang2008fundamentals} Even more than this, 
almost every aspect of our daily lives is permeated by interfacial 
phenomena.\cite{xing2019fluid} For this reason, an accurate description 
of fluid-solid interfaces is an invaluable tool in developing relevant 
scientific models and impactful technologies.

The purpose of this work is to examine the interfacial phenomenon 
of \textit{fluid slip}.\cite{navier1823memoire} This physical effect, 
quantified by the parameter known as \textit{slip length}, is usually 
neglected in macroscopic-scale fluid models and simulations when 
specifying velocity-field boundary conditions. Due to the slip length's 
typically small magnitude, often of the order of a few molecular diameters, 
the \textit{no-slip condition}\cite{day1990no, lauga2007microfluidics} is 
generally used as a suitable approximation for the fluid tangential velocity 
at a solid surface for most common applications.\cite{landau1987fluid, 
batchelor2000introduction}

However, the no-slip condition does have its shortcomings. In particular, for 
small-scale flows, such as those in micropores and nanotubes, the no-slip condition 
leads to significant errors in the fluid velocity field and, as a consequence, 
incorrect flow rates are obtained, sometimes underestimating observed values by 
several orders of magnitude.\cite{sokhan2002fluid, thomas2008reassessing, liu2011validity} 
Furthermore, an intrinsic limitation of the no-slip condition is that it is indifferent 
to changes in the interface materials since it is entirely independent of the molecular 
structure, atomic composition, and thermodynamic state of either fluid or solid surface. 
This disagrees with empirical data and prevents the use of theoretical models and 
numerical simulations in the design, development, or selection of functional interfaces 
for target applications.

Although the conventional theoretical framework of fluid mechanics, 
which relies on the \textit{continuum hypothesis}, is unable to 
determine interfacial boundary conditions, they can be readily 
assessed by employing atomistic simulation methods, such as 
\textit{molecular dynamics}.\cite{allen1987computer, rapaport2004art}
Moreover, in this way, \textit{material-specific} values of slip 
length can be computed for the system at hand and, additionally, 
their many parametric dependencies can be comprehensively analyzed.

Over the past few decades, plenty of effort has been directed towards 
understanding the mechanisms underlying fluid slip and the way it correlates 
with other physical quantities and phenomena.\cite{viscondi2019multiscale} 
In this regard, molecular-dynamics simulations have been employed in 
the investigation of fluid-slip processes\cite{lichter2004mechanisms, 
lichter2007liquid, martini2008molecular, yong2010investigating, 
sochi2011slip, wang2011slip, yong2013slip} and their connection 
with shear rate,\cite{thompson1997general, priezjev2007rate} 
channel size,\cite{xu2007boundary, ramos2016hydrodynamic} 
flow type,\cite{koplik1989molecular, cieplak2001boundary} 
wall stiffness,\cite{martini2008slip, asproulis2010boundary, 
asproulis2011wall, pahlavan2011effect} thermodynamic 
variables,\cite{guo2005temperature, servantie2008temperature, 
bao2017effects} wettability,\cite{barrat1999large, barrat1999influence, 
nagayama2004effects, voronov2006boundary, voronov2007slip, 
voronov2008review, huang2008water, sendner2009interfacial, 
ho2011liquid, huang2012friction, ramos2016wettability, 
yen2016effective} interatomic interactions,\cite{liu2009flow, 
liu2010surface, semiromi2010nanoscale, sofos2013parameters} 
as well as surface patterning and roughness.\cite{jabbarzadeh2000effect, 
cottin2003low, cottin2004dynamics, galea2004molecular, priezjev2005slip, 
priezjev2006influence, priezjev2007effect, niavarani2010modeling}

The present work particularly addresses the functional 
relationship between slip length and temperature. More 
specifically, we build upon the study carried out by 
Wang and Hadjiconstantinou,\cite{wang2019universal} 
where a reaction-rate model describing fluid slip as 
a \textit{thermally-activated process} was proposed. 
With the aid of non-equilibrium molecular-dynamics 
simulations, we evaluate the temperature dependence 
of slip length, as a function of the fluid-solid 
interaction strength, for an interface model consisting 
of a Lennard-Jones fluid and rigid walls. In this way, 
two completely opposite situations are observed, which 
we designate as \textit{slippery} and \textit{sticky} 
cases. The thermal-activation model is found to correctly 
describe both situations and precisely identify the 
interaction-parameter value at which the transition 
between them occurs. As shown in later sections, 
temperature is indeed a very influential variable 
on the slip-length value, with the ability to greatly 
amplify or suppress the fluid slip in the slippery 
and sticky cases, respectively.

The fundamental concepts and basic procedures 
employed in our numerical simulations and data 
analysis are briefly outlined in section~\ref{sec:Methodology}. 
Section~\ref{sec:Results} presents our main results, namely, 
an atomistic investigation of the fluid-slip thermal behavior 
for model interfaces. Further discussion on the results, based 
on the physical interpretation provided by the thermal-activation 
analytical model, is given in section~\ref{sec:Discussion}. 
Section~\ref{sec:Conclusion} comprises our concluding remarks 
and perspectives for future work. Details of the simulation 
setup are found in appendix~\ref{app:SimulationDetails}.

\section{Methodology}
\label{sec:Methodology}

\subsection{Slip Length}
\label{ssc:SlipLength}

The \textit{slip length}, denoted here by $\lambda$, is the parameter that
quantifies the fluid slip at a point~$\vec{r}$ of a solid surface~$\mcl{S}$ 
by relating the fluid-velocity tangential component~$u_{t}$, in the rest 
frame of $\mcl{S}$, to the local shear rate~$\dot{\gamma}=\del u_{t}/\del n$, 
that is,
\begin{equation}
	\left. u_{t}\right|_{\vec{r}\in\mcl{S}}
	=\lambda\left.\frac{\del u_{t}}{\del n}\right|_{\vec{r}\in\mcl{S}},
	\label{eq:SlipLength}
\end{equation}
\noindent where $n$ is the direction of the surface's normal pointing into 
the fluid. Geometrically, $\lambda$ is the distance one would have to linearly 
extrapolate the fluid velocity towards the solid surface so that $u_{t}$
vanishes. 

It is important to note that equation~\eqref{eq:SlipLength} is 
composed of \textit{effective} quantities, that is, both $u_{t}$ 
and its normal derivative must be understood as resulting from 
an extrapolation of the velocity field from some point sufficiently 
far into the fluid so that interfacial atomic-scale effects are not 
considered when evaluating $\lambda$. As a consequence, the slip length 
constitutes an \textit{apparent} quantity, whose value directly corresponds 
to the expected observation of a fluid-solid interface in a continuum-fluid 
perspective.

The applicability of definition~\eqref{eq:SlipLength} is twofold. In an atomistic simulation, 
as in the case of the results discussed in later sections, equation~\eqref{eq:SlipLength} 
is the working formula used to compute the slip length from statistically-evaluated 
fluid velocity fields. On the other hand, in the context of continuum fluid dynamics, 
expression~\eqref{eq:SlipLength} becomes a \textit{third-type boundary condition}, 
which can be used for solving the velocity field at fluid-solid interfaces, once 
slip-length values obtained by other means are provided.

\subsection{Simulations}
\label{ssc:Simulations}

The adopted interface model consists of a monoatomic fluid bounded 
by two solid walls whose atoms are kept in a rigid simple-cubic 
crystal arrangement. Fluid particles have mass~$m$ and both 
fluid-fluid and fluid-solid interactions are described by 
a pairwise \textit{Lennard-Jones potential},
\begin{equation}
	V(r)=4\vep\left[\left(\frac{\sigma}{r}\right)^{12}
	-\left(\frac{\sigma}{r}\right)^{6}\right],
	\label{eq:LennardJones}
\end{equation}
\noindent where $r$ is the distance between interacting particles. The 
parameter~$\vep$ controls the strength of the interaction, while $\sigma$ 
sets its length scale. For computational reasons, interactions are turned 
off for distances larger than a cutoff radius~$r_{c}$, so that forces are 
actually derived from a truncated and smoothed version of the potential, 
$\til{V}(r)=V(r)-V(r_{c})-\left.\del{V(r)}/\del{r}\right|_{r=r_{c}}(r-r_{c})$, 
thus ensuring the continuity of both energy and force at $r=r_{c}$.

Throughout this work, the values of $\sigma$ and $\vep$ determining the 
fluid-fluid interaction are considered, respectively, as the length and 
energy units. That is, a Lennard-Jones system of units is here employed 
with the parameters of the fluid-fluid potential set as reference. As a 
result, the temperature unit becomes $\vep/k_{B}$, where $k_{B}$ is the 
Boltzmann constant. In addition, by taking the mass~$m$ of a fluid particle 
as the unit of mass, it follows that the unit of time is $\tau=\sigma\sqrt{m/\vep}$. 
Note that, as a consequence of the adopted unit system, the Lennard-Jones 
parameters of the fluid-solid interaction, symbolized by $\vep_{fs}$ and 
$\sigma_{fs}$, have their values specified in relation to the fluid-fluid 
potential.

In order to extract slip-length values from an interface model, 
the fluid must be put in a shearing state. For this purpose, a 
non-equilibrium methodology based on a molecular-dynamics Couette 
flow is employed. The simulation box is made periodic in the $x$ 
and $y$ directions while solid walls constrain the fluid in the 
$z$~dimension. The walls, symmetrical and placed a distance~$h$ 
apart, move rigidly and uniformly in opposite directions along 
the $x$~direction with relative speed~$U$. In addition, the walls 
are made thick enough so that fluid particles are unaware of the 
solid's finite depth. After a transient period in its time evolution, 
the system reaches a steady state, which is characterized by a linear 
velocity profile~$u_{x}(z)$ in the fluid's bulk region, that is, in 
the volume portion sufficiently far from the solid walls. At this 
stage, the fluid also displays uniform bulk temperature and density 
profiles.

After ensuring that the system is in its steady state, the simulation 
enters in the production stage, in which velocities are sampled and 
averaged over several time steps. Once enough data is accumulated, 
the shear rate~$\dot{\gamma}$ is evaluated by applying a linear 
regression to the resulting bulk velocity profile. Finally, the 
slip length is obtained from equation~\eqref{eq:SlipLength}, which
yields the relation~$\lambda=(U/\dot{\gamma}-h)/2$ in the case of
a Couette flow.

All simulation runs are performed using the same number of particles 
and box dimensions. In particular, the wall separation distance~$h$ 
is kept constant so that the overall fluid density is fixed. However, 
as the fluid-solid interaction strength varies across runs, the 
layering effects at the interface cause slight variations in the 
bulk density.\footnote{The \textit{layering effect} is the well-known 
phenomenon of induced structuring of fluid molecules at the vicinity 
of a solid wall, causing oscillations of a characteristic wavelength 
in the fluid's density profile along the surface's normal direction. 
In our interface model, which is based on short-range Lennard-Jones 
interactions, these oscillations are observed only within a few 
molecular diameters away from the solid plates, so that the density 
is always uniform in the bulk region, where all measurements are 
made. Still, since different numbers of particles accumulate in 
the structured layers for different values of $\epsilon_{fs}$, 
and since the total number of fluid particles is the same for 
all simulations, the bulk density varies slightly across runs. 
As mentioned in the text, this variation in bulk density is 
negligible for our purposes.} These deviations are very 
small, though, and do not noticeably affect the fluid slip. 
Also, it should be mentioned that, since slip-length values 
are presumably connected with continuum boundary conditions, 
it is important to consider a sufficiently large simulation 
box, so that confinement effects, which would be spurious in 
this context, are avoided. Preliminary simulations have been 
performed in order to determine suitable box dimensions, 
for which the fluid slip is nearly invariant against further 
increases in the system size.

For the purposes of the intended analysis, each slip-length 
value must be determined at a well-defined temperature. Since 
the wall motion gives rise to a net inflow of energy, it is 
crucial to take measures to keep the temperature at a constant 
average value. Following previous works,\cite{thompson1997general,
pahlavan2011effect} a Langevin thermostat is applied to fluid 
particles. This procedure amounts to adding a combination 
of stochastic and dissipative forces to the system dynamics.
Given that $\vec{f}$ is the resultant conservative force 
acting on a fluid particle of acceleration~$\vec{a}$, 
velocity~$\vec{v}$, and position~$\vec{r}$, the 
thermostatted equations of motion read
\begin{equation}
	m\vec{a}=\vec{f}
	+\left\{-m\Gamma\left[\vec{v}-\vec{u}(\vec{r})\right]_{y}+S\right\}\hat{y},
	\label{eq:Langevin}
\end{equation}
\noindent where $\vec{u}$ is the fluid velocity field, $\Gamma$ is an adjustable 
damping coefficient, and $S(t)$ corresponds to a stochastic force satisfying 
$\bra S(t)\ket=0$ and $\bra S(t)S(t')\ket=2mT_{b}\Gamma\delta(t-t')$, with angle 
brackets denoting time averages. The imposed dynamics drive the system in such 
a way as to match the average fluid temperature~$T$ to the target value~$T_{b}$, 
representing the thermal-bath temperature. As shown in equation~\eqref{eq:Langevin}, 
only the particle acceleration in the $y$~direction is directly affected by `external' 
nonconservative forces. This is so in order to minimize the influence of the thermostat 
on the $x$ and $z$ directions, which are actually relevant to the Couette flow, thus 
reducing the effects of forces extraneous to the shear dynamics on the measured velocity 
profile.

In all performed simulations, the Langevin thermostat was able to adequately 
control the fluid's temperature, keeping its value spatially uniform and very 
close, except for temporal fluctuations, to the thermal bath's temperature 
(see the inset of figure~\ref{fig:Profiles}). In accordance, from now on, 
the symbol~$T$ is used here to denote the time-averaged temperature of 
the fluid, whose value is virtually identical to the target parameter~$T_{b}$.

According to previous studies,\cite{hansen1969phase, khrapak2010liquid, 
schultz2018comprehensive, stephan2019thermophysical} a low-density 
Lennard-Jones substance in thermodynamic equilibrium has a freezing
temperature close to $0.69\vep/k_{B}$. However, in the case of a 
non-equilibrium system, such as a Couette flow, the substance can
still present fluidic behavior at lower temperatures. Moreover, the 
actual freezing temperature~$T_{F}$ of a shearing fluid depends on 
its strength of interaction with the moving walls, as higher values
of the parameter~$\vep_{fs}$ facilitate the system's nucleation at 
the solid surfaces. For this reason, the temperature range used in 
our molecular-dynamics simulations differs for each choice of $\vep_{fs}$, 
starting at the lowest temperature for which the Lennard-Jones substance
properly displays a symmetrical Couette-flow pattern and going up 
to sufficiently high temperatures, where the slip-length variation
is slow.

A point deserving closer examination is the fact that, in the adopted 
methodology, the walls move rigidly, meaning that solid particles do not 
respond to the forces acting upon them. This is done mainly for practical 
reasons, since it makes simulations simpler to implement and computationally 
cheaper. Hence, solid-solid interactions play no role and, in particular, 
temperature effects on the walls are disregarded. The question then arises 
whether this simplification significantly affects the slip length's temperature 
dependence. A justification for the performed simulations can be found in 
previous studies,\cite{pahlavan2011effect} which have shown that, at least 
for simple wall models with sufficiently high stiffness, variations in the 
interaction strength between solid-surface particles have little impact 
on the slip length, provided the fluid temperature is properly kept 
constant.

Additional technical details of the performed simulations, such as 
the values of fixed parameters and the employed algorithm of time 
integration, are presented in appendix~\ref{app:SimulationDetails}.

\subsection{Slip model}
\label{ssc:SlipModel}

In their paper,\cite{wang2019universal} similarly to references 
[\onlinecite{lichter2007liquid}], [\onlinecite{martini2008molecular}],
and [\onlinecite{wang2011slip}], Wang and Hadjiconstantinou used an 
extension of Eyring's reaction-rate theory\cite{eyring1936viscosity, 
hirschfelder1964molecular, hanggi1990reaction} to develop a kinetic 
model\cite{blake1969kinetics, brochard1992dynamics} for the interfacial 
slip of simple fluids. For the purposes of the present work, their 
fluid-slip model can be summarized in the following formula:
\begin{equation}
	\lambda=\frac{\mu l^{2}}{\tau_{0}\Sigma k_{B}T}
	\exp\left(-\frac{V_{fs}+V_{ff}}{k_{B}T}\right),
	\label{eq:SlipLengthModel}
\end{equation}
\noindent where $T$ is the interface temperature and $\mu$ is the 
shear viscosity at the fluid's bulk region. The model is based on 
the idea that fluid particles in the first contact layer are driven 
by local shear forces and move by hopping over potential-energy barriers. 
Accordingly, the parameters $l$ and $\tau_{0}$ represent, respectively, 
the hop length and time scale. The energy barrier experienced by the 
fluid particles results from their interaction with both the solid wall 
and the fluid itself. This is expressed in equation~\eqref{eq:SlipLengthModel} 
by the quantities $V_{fs}$ and $V_{ff}$, which correspond respectively to the 
fluid-solid and fluid-fluid contributions to the barrier's height. As further 
explained in reference~[\onlinecite{wang2017molecular}], $\Sigma$ is the number 
of fluid particles per unit surface area in the first contact layer. Lastly, 
we should remark that equation~\eqref{eq:SlipLengthModel} represents the 
\textit{low-shear-rate limit} of a more general expression and, therefore, 
it is only valid when $\mu l\dot{\gamma}\ll 2\Sigma k_{B}T$.

The authors of reference~[\onlinecite{wang2019universal}] investigated 
the behavior of $\lambda$ as a function of several parameters involved 
in equation~\eqref{eq:SlipLengthModel}. As a result, excellent agreement 
was found between the model predictions and simulation data. Moreover, 
in the studied cases, it was observed that the influence of temperature 
on the slip-length value was very well described by the explicit dependence 
through the model's exponential factor alone. This is by no means obvious, 
since expression~\eqref{eq:SlipLengthModel} also has an explicit hyperbolic 
temperature factor and possibly many implicit temperature dependencies through 
the parameters $l$, $\tau_{0}$, $V_{fs}$, $V_{ff}$, and $\Sigma$. In view of 
these observations, following the aforementioned research, the slip length's 
temperature behavior is examined in the present work according to the simpler, 
scaling law
\begin{equation}
	\lambda=\lambda_{\infty}\exp{\left(-\frac{T_{c}}{T}\right)}.
	\label{eq:TrialFunction}
\end{equation}
\noindent In other words, expression~\eqref{eq:TrialFunction} 
is employed as a trial function in the regression analysis of 
the simulation results discussed in section~\ref{sec:Results}. 
Notice that the adjustable parameter~$\lambda_{\infty}$ corresponds 
to the high-temperature limiting value of the slip length, that 
is, $\lambda(T\gg |T_{c}|)\approx\lambda_{\infty}$. In its turn,
the \textit{characteristic temperature}~$T_{c}$ has a twofold
interpretation; its sign determines the overall thermal behavior 
of the slip length, whereas its absolute value establishes a 
temperature scale. 

By comparing equations \eqref{eq:SlipLengthModel} and \eqref{eq:TrialFunction}, note 
that the parameter~$T_{c}$ corresponds, except by a $k_{B}$~factor, to the total energy 
barrier~$V_{fs}+V_{ff}$. As mentioned in reference~[\onlinecite{wang2019universal}], in 
the case of pairwise Lennard-Jones interactions, the fluid-solid barrier contribution~$V_{fs}$ 
must result from a summation over terms with the form given by equation~\eqref{eq:LennardJones}.
As each of these terms is linearly proportional to $\vep_{fs}$, so is $V_{fs}$. Hence, by
writing $V_{fs}/k_{B}=\alpha\vep_{fs}$ and $V_{ff}/k_{B}=\beta$, the dependence of $T_{c}$ 
on the fluid-solid interaction strength can be made explicit:
\begin{equation}
	T_{c}=\alpha\vep_{fs}+\beta.
	\label{eq:TcAnsatz}
\end{equation}

The fact that the combination of physical quantities in the prefactor of
expression~\eqref{eq:SlipLengthModel} and the total energy barrier~$V_{fs}+V_{ff}$ 
seem to be mostly independent of temperature, giving rise to the parameters 
$\lambda_{\infty}$ and $T_{c}$ for a Lennard-Jones interface, is quite 
useful for the purposes of applying such a description to multiscale problems. 
Without this property, each component of equation~\eqref{eq:SlipLengthModel} 
would need to be determined as a function of the temperature, thus requiring 
a fair amount of atomistic simulations, just as if no model existed in the 
first place for the thermal behavior of the slip length. On the other hand, 
the use of equation~\eqref{eq:TrialFunction} is readily feasible in practical 
applications, since only a handful of simulations are required to properly 
evaluate $\lambda_{\infty}$ and $T_{c}$.

\section{Results}
\label{sec:Results}

\begin{figure}[hbt]
	\centering
	\includegraphics[width=0.45\textwidth]{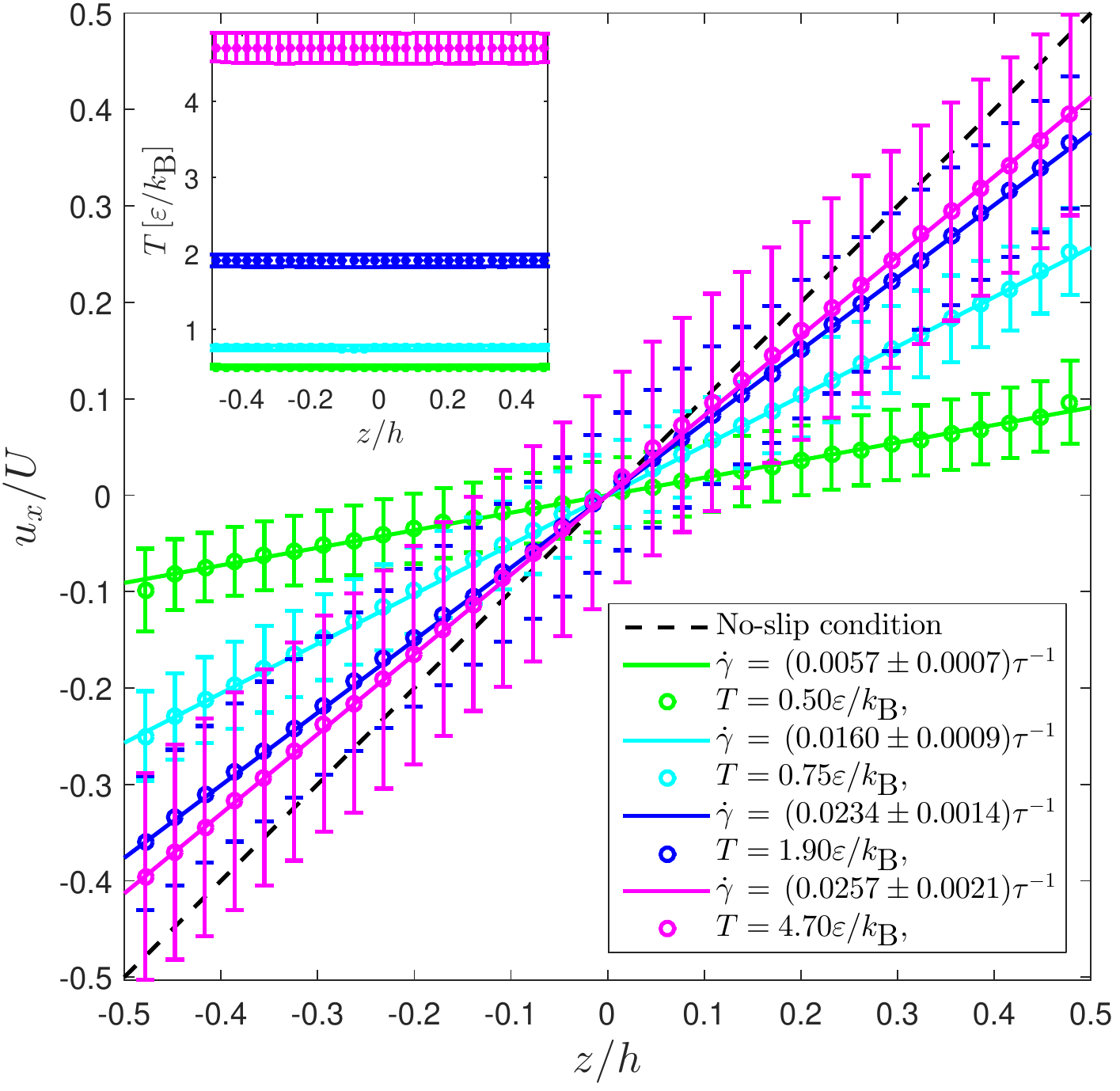}
	\vspace{-0.15cm}
	\caption{(Color online) Examples of simulation data sets 
	for a steady-state Couette-flow velocity profile, considering 
	$\vep_{fs}=0.2\vep$ and different values of $T$. Linear-regression 
	curves are also shown, as indicated by their respective shear-rate 
	values. The dashed line corresponds to the theoretical profile
	resulting from the no-slip condition, which is displayed here 
	as a reference. The figure inset shows the corresponding temperature 
	profiles, confirming that the average temperature is uniform across 
	the fluid and equal to the thermostat parameter~$T_{b}$.}
	\label{fig:Profiles}
\end{figure}

Figure~\ref{fig:Profiles} shows examples of steady-state velocity profiles 
obtained from atomistic Couette-flow simulations, considering fluid-solid 
interaction strength~$\vep_{fs}=0.2\vep$ and different values of fluid 
temperature~$T$. For comparison purposes, the curve corresponding to the 
no-slip condition is also displayed. As expected, in all presented cases, 
the velocity profiles are linear and symmetrical, except in the close 
vicinity of the solid surfaces, where small deviations from linearity 
are observed due to atomic-scale effects.

The shear rate~$\dot{\gamma}$ is readily obtained by calculating the 
slope of a Couette-flow velocity profile. Therefore, the values indicated 
in the legend of figure~\ref{fig:Profiles} were computed by performing 
linear regressions on the simulation results. Data points corresponding 
to atomic-scale interfacial nonlinearities were excluded from this fitting 
procedure, as only the fluid's bulk region must be considered in the shear-rate 
assessment. Errors associated with $\dot{\gamma}$ were evaluated by propagating 
the standard deviations of the velocity profile, in accordance with the usual 
formulas from basic error analysis.\cite{taylor1997introduction, bevington2003data}

As evidenced by figure~\ref{fig:Profiles}, in the case of $\vep_{fs}=0.2\vep$, 
the shear rate increases with the fluid temperature. Also, note that the 
velocity profile approaches the situation predicted by the no-slip condition 
for higher values of $T$. This behavior is opposite to that observed by Wang 
and Hadjiconstantinou,\cite{wang2019universal} as only cases corresponding 
to $T_{c}>0$ in equation~\eqref{eq:TrialFunction}, implying a monotonically 
decreasing shear rate, were reported in their work. In our investigation, 
we also contemplate the possibility of the parameter~$T_{c}$ becoming negative, 
which gives rise to an exponential increase of the slip length as the interface 
temperature is reduced. Notice that this situation is particularly interesting, 
as $\lambda$ can reach arbitrarily large values at low temperatures, resulting 
in the complete deviation from the no-slip condition, possibly even at macroscopic 
scales. In practice, however, the slip-length value does have an effective upper 
bound for negative $T_{c}$, which is attained at the threshold of solidification, 
as the concept of slip length loses its meaning below the fluid's freezing 
temperature~$T_{F}$.

\begin{figure}[htb]
	\centering
	\includegraphics[width=0.45\textwidth]{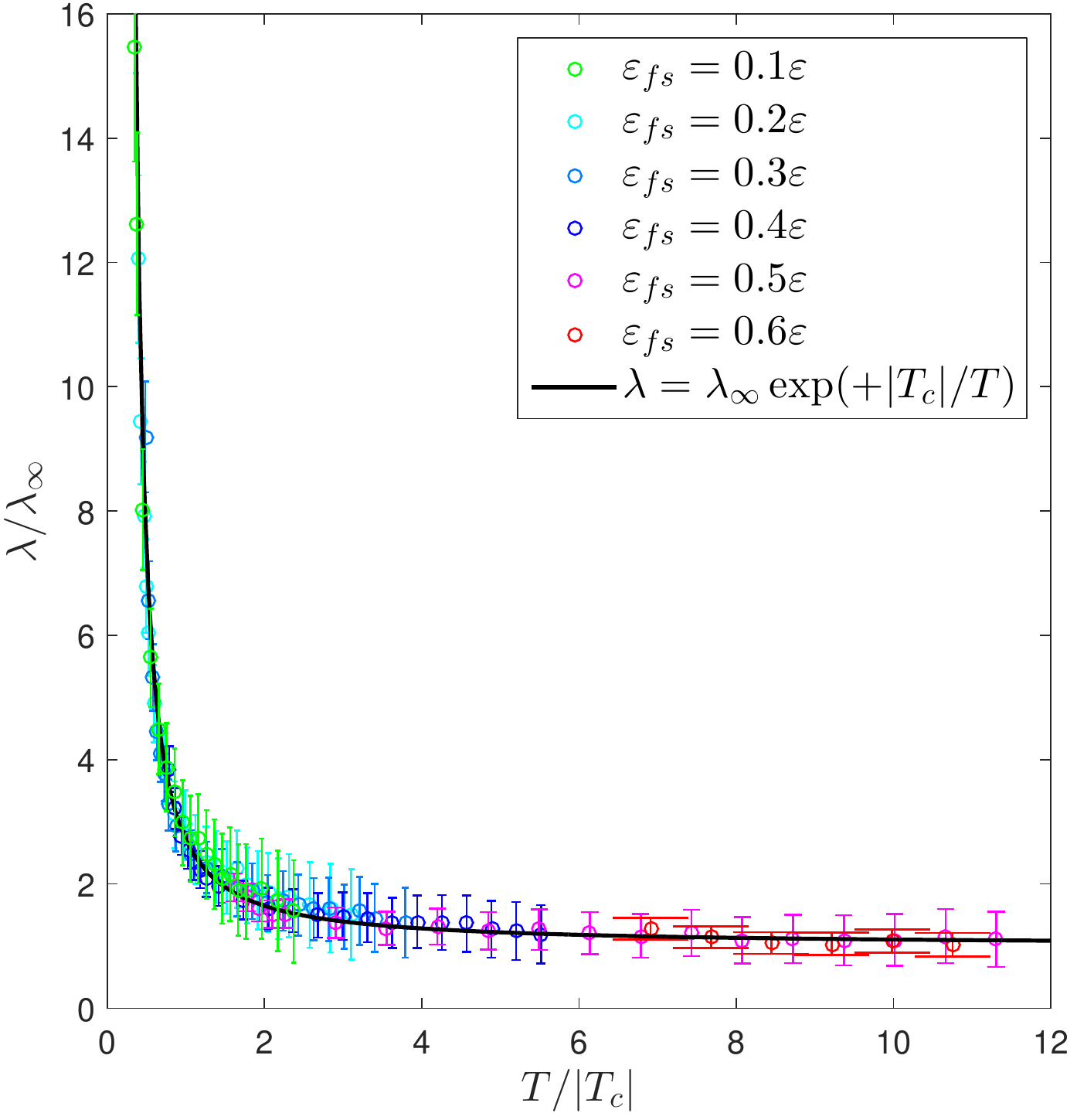}
	\caption{(Color online) Slip length as a function of 
	the fluid temperature for several values of $\vep_{fs}$ 
	resulting in a slippery behavior ($T_{c}<0$).}
	\label{fig:SlipperyCase}
\end{figure}

Figure~\ref{fig:SlipperyCase} presents the relationship between 
the slip length and the fluid temperature for different values 
of the parameter~$\varepsilon_{fs}$, considering only cases in 
which $T_{c}<0$. This regime, characterized by an upper-unbounded
and monotonically-decreasing curve~$\lambda(T)$, is designated 
as the \textit{slippery thermal behavior}. Each displayed value 
of $\lambda$ resulted from a single Couette-flow simulation and
its error bar was obtained by propagating the associated error 
found for $\dot{\gamma}$.

Note that, in figure~\ref{fig:SlipperyCase}, $\lambda$ and $T$ 
are normalized by their corresponding values of $\lambda_{\infty}$ 
and $|T_{c}|$. In this way, all data sets can be compared to a single 
reference curve, indicating a universal behavior shared by different 
interfaces. The parameters $\lambda_{\infty}$ and $T_{c}$ were 
determined by performing linear regressions with the logarithm 
of expression~\eqref{eq:TrialFunction} as trial function, 
resulting in the values presented by table~\ref{tb:Regression}.

\begin{table}[htb]
	\caption{Regression results for model function~\eqref{eq:TrialFunction}.}
	\label{tb:Regression}
	\begin{tabular}{c c c c}
		$\vep_{fs}\;[\vep]$ 
		& $\lambda_{\infty}\;[\sigma]$ 
		& $T_{c}\;[\vep/k_{B}]$ 
		& $R^{2}$ \\[1pt] \hline
		$0.1$ & $1.9\pm0.2$ & $-1.98\pm0.09$ & $0.97$ \\ \hline
		$0.2$ & $2.2\pm0.2$ & $-1.51\pm0.07$ & $0.96$ \\ \hline
		$0.3$ & $2.7\pm0.3$ & $-1.03\pm0.07$ & $0.97$ \\ \hline
		$0.4$ & $3.0\pm0.3$ & $-0.63\pm0.07$ & $0.98$ \\ \hline
		$0.5$ & $3.2\pm0.4$ & $-0.31\pm0.08$ & $0.96$ \\ \hline
		$0.6$ & $3.3\pm0.4$ & $-0.07\pm0.08$ & $0.28$ \\ \hline
		$0.7$ & $3.4\pm0.5$ & $+0.24\pm0.12$ & $0.97$ \\ \hline
		$0.8$ & $3.6\pm0.6$ & $+0.53\pm0.18$ & $0.98$ \\ \hline
		$0.9$ & $3.5\pm0.7$ & $+0.76\pm0.26$ & $0.98$ \\ \hline
		$1.0$ & $3.5\pm0.7$ & $+0.94\pm0.26$ & $0.97$ \\ \hline						
		$1.1$ & $3.7\pm1.1$ & $+1.33\pm0.44$ & $0.98$ \\ \hline
		$1.2$ & $3.7\pm1.4$ & $+1.48\pm0.72$ & $0.98$ \\ \hline		
	\end{tabular}
\end{table}

The coefficient of determination~$R^{2}$, also shown in table~\ref{tb:Regression}, 
confirms that equation~\eqref{eq:TrialFunction} provides a suitable description 
for the functional relationship between slip length and temperature in the 
case of slippery interfaces. Observe that the values of $R^{2}$ are close 
to unity, evidencing the good agreement of the analytical model with the 
simulation data, except for $\vep_{fs}=0.6$. However, as can be visually 
verified in figure~\ref{fig:SlipperyCase}, function~\eqref{eq:TrialFunction}
does correctly predict the interface's temperature dependence 
for all values of the parameter~$\vep_{fs}$.

In the critical situation of $\vep_{fs}=0.6$, according to the simulation 
results, the slip length is practically invariant with respect to temperature. 
This fact is reflected in the almost null value of the parameter~$T_{c}$, 
which in turn implies that expression~\eqref{eq:TrialFunction} is approximately 
constant. By definition, the coefficient~$R^{2}$ provides a quantification 
of fit quality by comparing the proposed model function to the average 
value of the dependent variable. Therefore, in the case where the trial 
curve corresponds to a constant, no conclusions can be drawn from $R^{2}$,
thus explaining the coefficient's small value for $\vep_{fs}=0.6$.

\begin{figure}[htb]
	\centering
	\includegraphics[width=0.45\textwidth]{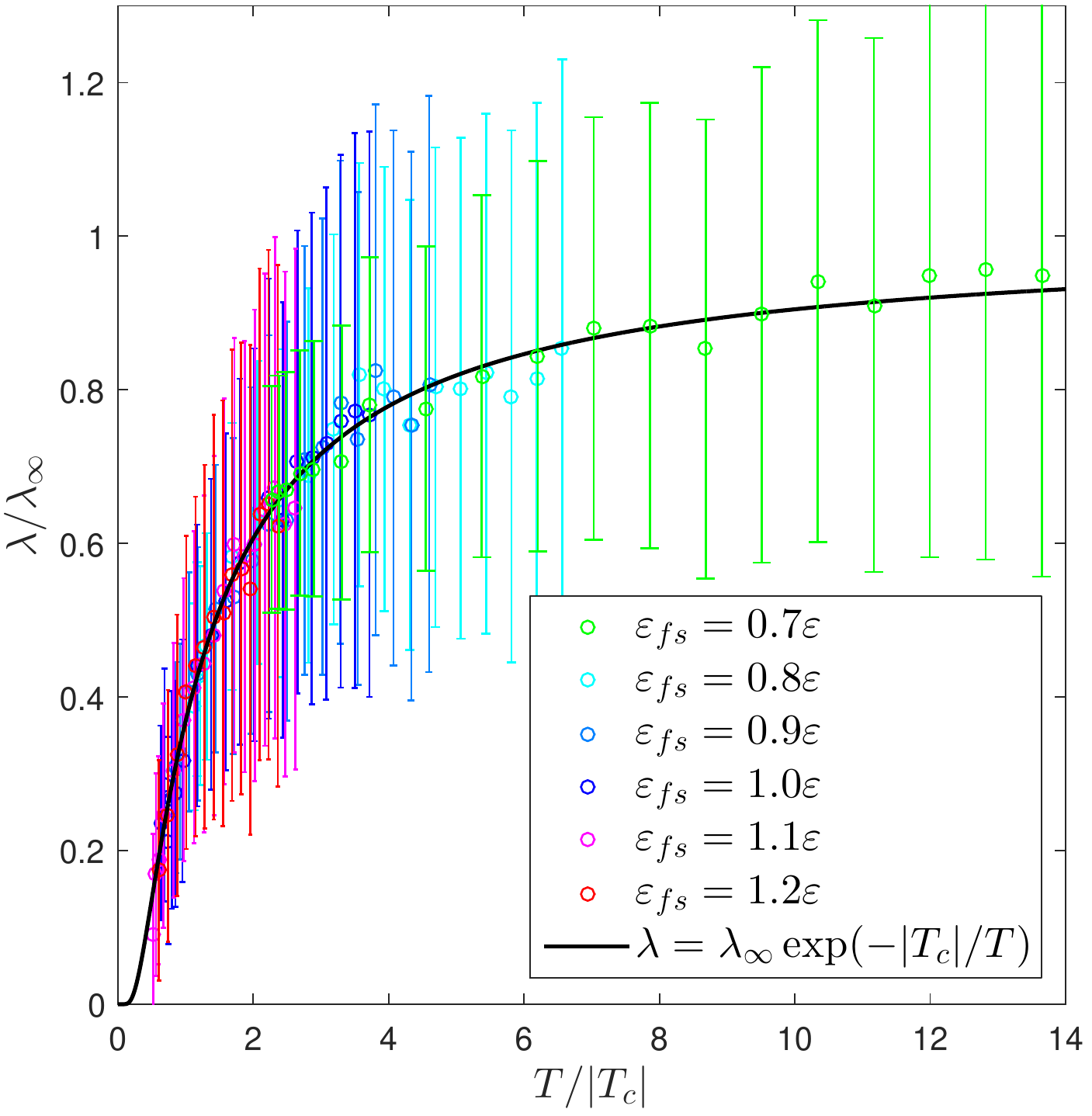}
	\caption{(Color online) Slip-length values as a function of 
	the fluid temperature for several values of $\vep_{fs}$ 
	resulting in a sticky behavior ($T_{c}>0$).}
	\label{fig:StickyCase}
\end{figure}

Simulation data portraying the slip length as a function of the fluid 
temperature is again shown in figure~\ref{fig:StickyCase}, now considering 
values of the parameter~$\vep_{fs}$ corresponding to $T_{c}>0$. Similarly 
to the preceding case, the variables $\lambda$ and $T$ were normalized 
by their respective values of $\lambda_{\infty}$ and $|T_{c}|$, allowing 
comparison with a single reference curve. Obtained values of $\lambda_{\infty}$, 
$T_{c}$, and $R^{2}$, resulting from linear regressions using the logarithm 
of function~\eqref{eq:TrialFunction}, are also found in table~\ref{tb:Regression}.

As can be visually inferred from figure~\ref{fig:StickyCase} and confirmed 
by the close-to-unity values of $R^{2}$, expression~\eqref{eq:TrialFunction}
appropriately describes the temperature dependence of interfacial slip also 
for positive $T_{c}$. In this situation, the model function~$\lambda(T)$ 
is monotonically increasing and bounded between zero and $\lambda_{\infty}$.
As a consequence, if two distinct hypothetical interfaces are considered, 
with identical values of $\lambda_{\infty}$ and $|T_{c}|$, but opposite 
signs of the characteristic temperature, the slip length of the interface 
with positive $T_{c}$ will be strictly lower for any finite fluid temperature. 
The difference between the two cases is even more evident for low temperatures
($T\ll|T_{c}|$), as $\lambda$ approaches zero for $T_{c}>0$, while it goes 
to infinity for $T_{c}<0$. For these reasons, the regime of positive $T_{c}$ 
is designated as the \textit{sticky thermal behavior}.

According to table~\ref{tb:Regression}, in both slippery and sticky cases, 
the value of $\lambda_{\infty}$ slowly increases as the parameter~$\vep_{fs}$ 
is raised. The increase rate is different for each of the thermal behaviors, 
being clearly faster in the slippery regime. Therefore, albeit counterintuitive, 
simulation data does indicate that stronger interface interactions result in 
larger slip lengths at high fluid temperatures ($T\gg|T_{c}|$).

\begin{figure}[htb]
	\centering
	\includegraphics[width=0.45\textwidth]{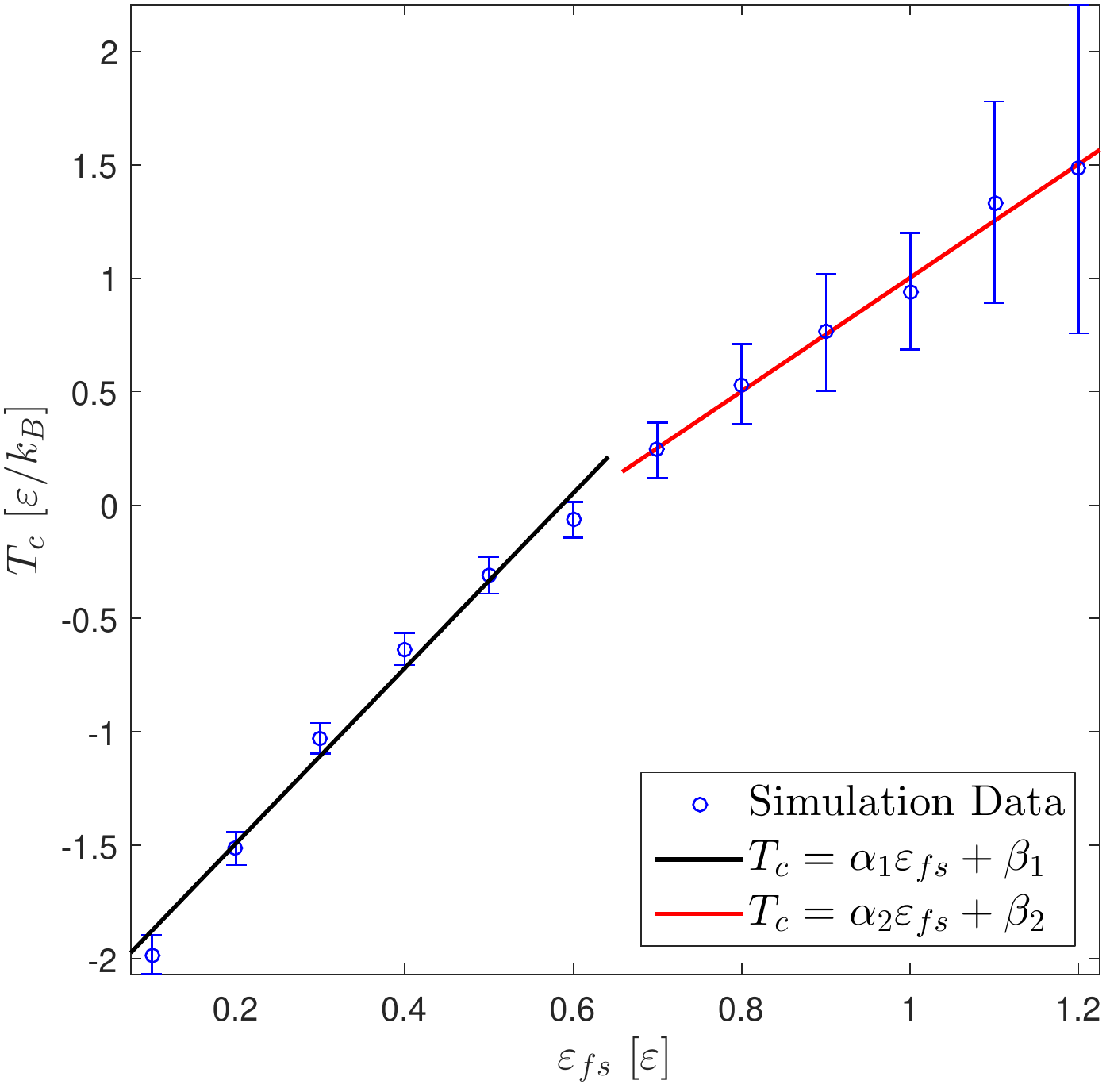}
	\caption{(Color online) Dependence of $T_{c}$ on the fluid-solid 
	interaction parameter~$\vep_{fs}$.}
	\label{fig:TcCurve}
\end{figure}

As expected from equation~\eqref{eq:TcAnsatz}, the parameter~$T_{c}$ linearly increases 
with the interaction strength~$\vep_{fs}$. However, as shown in figure~\ref{fig:TcCurve}, 
different values of the adjustable coefficients $\alpha$ and $\beta$ are found for each 
thermal behavior, once again indicating profound changes in the interfacial properties 
as the slippery-sticky transition takes place. For $T_{c}<0$, a linear-regression analysis 
provides $\alpha_{1}=(3.9\pm0.2)k_{B}^{-1}$ and $\beta_{1}=(-2.26\pm0.07)\vep/k_{B}$ 
with coefficient of determination~$R^{2}=0.984$. Similarly, for $T_{c}>0$, the values 
$\alpha_{2}=(2.5\pm0.7)k_{B}^{-1}$ and $\beta_{2}=(-1.50\pm0.55)\vep/k_{B}$ are obtained, 
with $R^{2}=0.991$. Notice that, in both cases, high agreement between the model function
and the numerical simulations is evinced by the coefficient~$R^{2}$. 

By employing the fitted curves of figure~\ref{fig:TcCurve}, the value 
of the parameter~$\vep_{fs}$ at which the slippery-stick transition occurs 
can be readily determined. This critical fluid-solid interaction strength, 
denoted by $\til{\vep}_{fs}$, directly follows from equation~\eqref{eq:TcAnsatz}
by setting $T_{c}=0$, that is, $\til{\vep}_{fs}=-\beta/\alpha$. This relation 
yields $\til{\vep}_{fs,1}=(0.59\pm0.03)\vep$, for negative $T_{c}$, and 
$\til{\vep}_{fs,2}=(0.60\pm0.27)\vep$, for positive $T_{c}$. Note that 
the values of $\til{\vep}_{fs,1}$ and $\til{\vep}_{fs,2}$ agree within 
errors, as required for consistency and continuity, since both curves 
for $T_{c}(\vep_{fs})$ are expected to meet at the critical parameter. 
In addition, the two estimates for $\til{\vep}_{fs}$ suggest that 
its true value is close to $0.6\vep$, as could be inferred from 
figure~\ref{fig:SlipperyCase}, due to the approximately constant 
behavior of $\lambda(T)$ at this interaction strength.

\section{Discussion}
\label{sec:Discussion}

As shown in the previous section, both slippery and sticky thermal behaviors 
were adequately described by a previously-proposed model function,\cite{wang2019universal} 
which was linearly fitted to our simulation data, thus determining the adjustable 
parameters $\lambda_{\infty}$ and $T_{c}$. It is particularly interesting to note 
that this analytical model also correctly portrays the slippery regime, although 
its original authors seem to have devised it with only the sticky case in mind. 
As mentioned earlier, in the derivation outlined by Wang and Hadjiconstantinou, 
the fluid-slip phenomenon is understood as a thermally-activated hopping process, 
where the thermal energy enables fluid particles to overcome potential barriers. 
Therefore, in this context, the existence of interfaces for which slip length 
decreases with temperature appears to be counterintuitive.

Within the slippery regime, the characteristic temperature~$T_{c}$ is negative, 
meaning that the model's total energy barrier~$V_{fs}+V_{ff}=k_{B}T_{c}$ is 
actually a potential well, promoting the movement of fluid particles along 
the solid surface. This slip-facilitating mechanism is gradually suppressed 
by increasing the thermal fluctuations, as they may cause the entrapment of 
fluid particles in the potential wells.

In general, interfacial fluid particles are subjected to a competition between 
cohesive and adhesive forces. Seeking to maintain cohesion, forces originating 
from the fluid are responsible for the term~$V_{ff}=k_{B}\beta$, which is always 
negative, as follows from the values of $\beta_{1}$ and $\beta_{2}$ presented 
in section~\ref{sec:Results}. On the other hand, the adhesive forces from 
the solid surface result in the term~$V_{fs}=k_{B}\alpha\vep_{fs}$, which 
is strictly positive, as evinced by the obtained values of $\alpha_{1}$ 
and $\alpha_{2}$. Therefore, the slippery behavior occurs when the fluid 
cohesion at the interface is stronger than the adhesion to the solid 
surface ($|V_{ff}|>|V_{fs}|$). In the opposite situation, the sticky 
behavior arises.

The \textit{slippery-sticky transition} takes place at the critical 
interaction-strength value~$\til{\vep}_{fs}$, for which the characteristic 
temperature vanishes or, equivalently, the energy contributions $V_{fs}$ 
and $V_{ff}$ cancel each other out. In this particular situation, fluid 
particles experience a smooth average energy landscape at the interface, 
so that thermal mechanisms of hopping and entrapment are absent, resulting 
in the slip length's temperature invariance. Consequently, at the transition, 
the slip-length value is solely determined by other physical properties, 
such as viscosity and surface density.

In section~\ref{sec:Results}, by examining the dependence of $T_{c}$ on the 
fluid-solid interaction strength, the value of $\til{\vep}_{fs}$ was determined 
to be approximately $0.6\vep$. Interestingly, in addition to its sign change, 
the curve~$T_{c}(\vep_{fs})$ also exhibits a slope discontinuity at the transition. 
Therefore, the characteristic temperature signals the slippery-sticky transition 
in two different ways, establishing itself as a reliable identifier parameter 
for the slip length's thermal behavior.\footnote{Other potentially relevant 
interface properties, such as the fluid-solid radial distribution function 
and fluid's density profile close to the solid surface, were also examined 
as functions of temperature for several values of the fluid-solid interaction 
strength. No remarkably discriminating behavior was observed in these quantities 
when comparing slippery and sticky interfaces. In general, these observables 
change continuously with $\vep_{fs}$, showing no evident sign of the transition 
at $\til{\vep}_{fs}$. For this reason, we chose not to show these negative results 
here.}

Finally, we would like to point out the relevance of $|T_{c}|$ to the 
interfacial thermal behavior, particularly when compared to the fluid's 
freezing temperature~$T_{F}$, since the latter represents the lower bound 
of the temperature range in which the present analysis is valid.
As previously discussed, for small absolute values of the characteristic 
temperature ($|T_{c}|\ll T_{F}$), the slip length is almost independent 
of $T$. In general, for $|T_{c}|$ below the fluid's freezing temperature~$T_{F}$, 
$\lambda(T)$ is a slowly varying function within the accessible temperature 
range. On the other hand, for $|T_{c}|>T_{F}$, $\lambda(T)$ can present 
actual exponential behavior for sufficiently low temperatures ($T\ll|T_{c}|$), 
quickly bringing the slip-length value to zero or taking it to infinite, 
depending on the sign of $T_{c}$, as can be observed in figures 
\ref{fig:SlipperyCase} and \ref{fig:StickyCase}. Therefore, if 
a technological application demands an exceptionally slippery 
or sticky interface, the interacting materials should be chosen 
in such a way as to provide the highest possible value of $|T_{c}|$, 
so that the system's working temperature is comparatively low and, 
consequently, an appropriate slip length is attainable.

\section{Conclusion}
\label{sec:Conclusion}

By performing molecular-dynamics simulations of a Couette flow, composed 
of a Lennard-Jones fluid at constant density and simple-cubic rigid walls, 
the temperature effects on fluid slip were thoroughly examined. As shown 
in section~\ref{sec:Results}, the slip length can present two completely 
different thermal behaviors depending on the interaction strength between 
the fluid and the solid surfaces. In the \textit{slippery case}, corresponding 
to weak interface interactions ($\vep_{fs}<\til{\vep}_{fs}$), the slip-length 
value is upper-unbounded and decreases with temperature. On the other hand, 
for strong fluid-solid interactions ($\vep_{fs}>\til{\vep}_{fs}$), the 
interface displays the \textit{sticky behavior}, characterized by slip-length 
values increasing with temperature, but in a restricted interval. 

From a technological point of view, interfaces capable of operating in the 
slippery regime are very promising. Assuming that the slippery thermal behavior 
is not hindered by other effects, the arbitrarily large slip lengths obtainable 
at low temperatures ($T\ll|T_{c}|$) could be readily exploited in applications 
requiring low frictional stress. From a design perspective, the determination 
of characteristic temperatures, either empirically or numerically, provides 
a searching path for the development or selection of materials for such 
friction-reduction purposes. As previously discussed, large-magnitude negative 
values of $T_{c}$ will make the slippery behavior more appreciable at higher 
temperatures, ideally far above the fluid's solidification temperature~$T_{F}$.

The present work's findings, in conjunction with the analytical model proposed 
in reference~[\onlinecite{wang2019universal}], can assist in devising methods 
for the efficient simulation of continuum fluidic systems in which interfacial 
boundary conditions are required to be sensitive to nonuniform temperature 
distributions. By performing a small set of molecular-dynamics simulations, 
the parameters $\lambda_{\infty}$ and $T_{c}$ can be promptly determined 
for a fluid-solid interface of interest. Upon substituting these values 
into the slip-length model function~\eqref{eq:TrialFunction} and then 
inserting the resulting expression into equation~\eqref{eq:SlipLength}, 
the velocity-field boundary conditions at interfaces displaying 
spatial or temporal temperature variations can be readily 
specified, giving rise to a \textit{hierarchical multiscale 
approach} to fluid dynamics.\cite{viscondi2019multiscale}

Clearly, the physical interpretation of our results, as laid 
out in section~\ref{sec:Results}, strongly relies on the model 
proposed in reference~[\onlinecite{wang2019universal}]. Despite 
this, the present work should not be viewed as an attempt to 
access the microscopic validity of Wang and Hadjiconstantinou's 
model, that is, whether the thermal hopping and entrapment mechanisms, 
besides providing a useful conceptualization of the slip phenomenon, 
actually take place in reality. Rather, our core result is the observation 
that, concerning the slip phenomenon in a simple interface system, 
two distinct thermal behaviors can occur depending on the fluid-solid 
interaction parameter, an observation purely based on numerical simulations. 
The reported data, nonetheless, provide new evidence in favor of the model, 
namely, the particularly compelling fact that it is capable of correctly 
describing both observed behaviors while also endowing the results with 
an insightful explanation that fits well within the reaction-rate picture, 
as discussed in section~\ref{sec:Discussion}. The investigation of the 
analytical model's microscopic realism is a very interesting topic, which 
we will hopefully address in future work.

Finally, note that the results of section~\ref{sec:Results} are reasonably 
general. That is, by adopting a system of units based on the fluid's fundamental 
properties and defining the fluid-solid interaction parameters relative to it, 
a representative discussion is provided for any interface whose intermolecular 
forces can be effectively described by Lennard-Jones potentials. However, for more
complex interparticle interactions, other effects beyond those considered in 
the present study could influence the slip length's temperature dependence, 
so that it would be erroneous to conclude that the slippery and sticky behaviors, 
as described by function~\eqref{eq:TrialFunction} and portrayed by figures 
\ref{fig:SlipperyCase} and \ref{fig:StickyCase}, will be observed in more 
sophisticated interface models. Nevertheless, even in more complicated 
scenarios, the mechanisms investigated in this work still participate in 
the system dynamics, and they will have a role in determining the interfacial 
fluid slip. This is because the repulsive and van der Waals interactions 
following from the Lennard-Jones potential are essential ingredients of more 
elaborated force fields, which will typically also include long-range Coulomb 
forces as well as other terms that account for the internal degrees of freedom 
of each molecule. Whether the Lennard-Jones contribution has a generally dominant 
or negligible role in determining the slip length's thermal behavior in such 
complex systems remains an open question that we intend to tackle 
in future studies.

\section*{Acknowledgments}

This work is part of a project developed in the RCGI (Research Centre for Gas Innovation), 
with support from Shell, CNPq (National Council for Scientific and Technological Development), 
under grant number 2302665/2017-0, and FAPESP (S\~ao Paulo Research Foundation), under grant 
numbers 2014/50279-4, 2018/03211-6, and 2020/50230-5.

\section*{Data Availability}

The data that support the findings of this study are available from the corresponding 
author upon reasonable request.

\appendix

\section{Simulation Details}
\label{app:SimulationDetails}

This appendix presents further details on the adopted methodology
and employed algorithms. In particular, the parameter values which 
remain fixed across all numerical simulations in section~\ref{sec:Results} 
are provided here. 

Firstly, we point out that an in-house code was used for performing 
the molecular-dynamics simulations. A representative subset of the 
obtained data was also compared to its equivalent counterpart generated 
with the aid of the \textit{LAMMPS} package,\cite{plimpton1995fast} 
yielding similar results within error bars.

As mentioned in subsection~\ref{ssc:Simulations}, the fluid particles 
interact with one another through the truncated and smoothed Lennard-Jones 
potential, with its parameters $\sigma$ and $\vep$ serving as a reference
for the system of units. In addition, the cutoff radius of fluid-fluid 
interactions receives the customary value~$r_{c}^{ff}=2.5\sigma$. In all 
considered simulations, the fluid is composed of $12000$ particles, which 
occupy a channel of width~$h=32.18\sigma$, specifically chosen to avoid 
the occurrence of confinement effects, as determined by early investigations.
The $x$ and $y$ directions of the simulation box, in which periodic boundary 
conditions are assumed, have the same length~$L=21.45\sigma$. Considering 
the channel's available volume~$\mcl{V}_{f}=hL^{2}$, notice that the fluid 
has average density~$\rho_{f}=0.81\sigma^{-3}$. Additionally, in order to 
properly control the fluid temperature, equation~\eqref{eq:Langevin} is 
employed with damping coefficient~$\Gamma=1.0\tau^{-1}$.

The planar Couette flow is generated by two flat plates moving in parallel 
with relative speed~$U=1.0\sigma/\tau$. These rigid walls have simple-cubic 
crystal structure, with their $(001)$~planes facing the fluid and their atoms 
aligned parallel to the specified Cartesian directions. Each solid surface 
is composed of $6936$ particles, arranged in a layer of thickness~$h_{s}=3.78\sigma$ 
and average density~$\rho_{s}=4.0\sigma^{-3}$. The atomic structure, density, 
and velocity of the plates were chosen in order to accentuate the fluid-slip 
phenomenon, in agreement with preliminary simulations. These earlier 
investigations have shown that increasing the shear rate, by raising 
the plate velocities, implies larger slip-length values, as intuitively 
expected and demonstrated by other authors.\cite{thompson1997general, priezjev2007rate} 
Interfacial slip was also found to occur more intensely for solids with higher particle
densities, in accordance with the idea that a more compact crystal structure displays 
a smoother surface thus impairing tangential momentum transfer. Additionally, the 
simple-cubic structure was found to yield the largest slip lengths among the cubic 
lattices.

The solid-surface particles are restricted to uniform linear motion, with 
velocity~$+U\hat{x}/2$ in the upper plate and $-U\hat{x}/2$ in the lower 
one. Accordingly, these particles are not subject to forces from the fluid 
or the solid surface itself. Although there is no reaction, wall atoms act 
on the fluid through the truncated and smoothed Lennard-Jones potential with 
fixed parameters $\sigma_{fs}=0.75\sigma$ and $r_{c}^{fs}=2.5\sigma_{fs}$. 
Note that the range of the fluid-solid interaction is appropriately smaller 
than the plate thickness~$h_{s}$, so that the fluid particles are unable to 
`perceive' the solid surfaces as finite objects.

The numerical integration of the dynamical system, composed of 
coupled Langevin equations, is performed by a leapfrog integrator, 
considering fixed time steps of length~$\Delta t=5\times 10^{-3}\tau$. 
The Couette-flow simulation is divided into two stages. In a first 
phase, lasting~$\mcl{T}=5\times10^{5}\Delta t=2.5\times 10^{3}\tau$, 
the fluid, initially at rest, is sheared by the solid surfaces until 
the steady state is reached. As initial atomic positions, fluid particles 
are positioned in a regular lattice with random-direction velocities, whose 
magnitudes are chosen so that each atom's kinetic energy conforms to the 
target temperature~$T_{b}$. In a second phase, also of duration~$\mcl{T}$, 
the values of all pertinent physical quantities are collected at each 
time step. 

During the data-gathering stage, the simulation box is 
divided along the $z$~direction into $40$ slabs of identical 
thickness. In this way, instantaneous values of the fluid's 
density, velocity, and temperature profiles are evaluated, 
considering respectively the number, the mean velocity, 
and the mean thermal-kinetic energy of the particles in 
each partition. Time averages and their respective standard 
deviations are then calculated by taking into account all 
time steps in the period~$\mcl{T}$. Note that, according 
to this procedure, the standard deviations are proportional
to the fluctuation amplitudes of the physical quantities
during their time series.

In order to properly investigate the slip length's temperature dependence, 
molecular-dynamics simulations were performed with temperature values between 
$0.1\vep/k_{B}$ and $3.5\vep/k_{B}$ for each choice of the parameter~$\vep_{fs}$. 
Since part of this range lies below the Lennard-Jones equilibrium freezing 
temperature, approximately $0.69\varepsilon/k_{B}$, some of the low-temperature 
runs naturally resulted in systems displaying no fluidic behavior and, consequently, 
were discarded from the fluid-slip analysis. In other words, only those simulations 
presenting a steady-state characterized by a linear and symmetrical bulk velocity 
profile, as well as uniform bulk temperature and density profiles, after the 
period~$\mcl{T}$ of system preparation, were effectively considered in the 
slip-length evaluation.

Linear regressions, employed in the determination of shear-rate values 
and the adjustable parameters in expressions \eqref{eq:TrialFunction}
and \eqref{eq:TcAnsatz}, were performed with the method of least squares, 
considering the inverse squared errors of the dependent variables as 
summation weights.\cite{bevington2003data}


\end{document}